\documentclass[article,reprint,amsmath,amssymb,superscriptaddress,longbibliography]{revtex4-1}

\usepackage[colorlinks,
linkcolor=blue,
anchorcolor=blue,
citecolor=blue,
]{hyperref}

\usepackage{graphicx}
\usepackage{epstopdf}
\usepackage{hyperref}
\UseRawInputEncoding
\setcitestyle{super} 

\usepackage{breakurl}
\usepackage{multirow}
\usepackage{enumitem}
\usepackage{color}

\usepackage{tablefootnote}
\usepackage{threeparttable}
\usepackage{tabularx}
\usepackage{booktabs}
\usepackage{textcomp}
\usepackage{lineno}

\begin{document}

\title {Superconducting diode effect and interference patterns in Kagome CsV$_3$Sb$_5$}

\author{Tian Le}
\thanks{Equal contributions} 
\affiliation{Key Laboratory for Quantum Materials of Zhejiang Province, Department of Physics, School of Science and Research Center for Industries of the Future, Westlake University, Hangzhou 310030, P. R. China} 
\affiliation{Institute of Natural Sciences, Westlake Institute for Advanced Study, Hangzhou 310024, P. R. China}

\author{Zhiming Pan}
\thanks{Equal contributions}
\affiliation{Key Laboratory for Quantum Materials of Zhejiang Province, Department of Physics, School of Science and Research Center for Industries of the Future, Westlake University, Hangzhou 310030, P. R. China} 
\affiliation{Institute of Natural Sciences, Westlake Institute for Advanced Study, Hangzhou 310024, P. R. China}
\affiliation{Institute for Theoretical Sciences, Westlake University, 310024, Hangzhou, China}

\author{Zhuokai Xu}
\affiliation{Key Laboratory for Quantum Materials of Zhejiang Province, Department of Physics, School of Science and Research Center for Industries of the Future, Westlake University, Hangzhou 310030, P. R. China} 
\affiliation{Institute of Natural Sciences, Westlake Institute for Advanced Study, Hangzhou 310024, P. R. China}

\author{Jinjin Liu}
\affiliation{Centre for Quantum Physics, Key Laboratory of Advanced Optoelectronic Quantum Architecture and Measurement (MOE), School of Physics, Beijing Institute of Technology, Beijing 100081, China} 
\affiliation{Beijing Key Lab of Nanophotonics and Ultrafine Optoelectronic Systems, Beijing Institute of Technology, Beijing 100081, China}

\author{Jialu Wang}
\affiliation{Key Laboratory for Quantum Materials of Zhejiang Province, Department of Physics, School of Science and Research Center for Industries of the Future, Westlake University, Hangzhou 310030, P. R. China} 
\affiliation{Institute of Natural Sciences, Westlake Institute for Advanced Study, Hangzhou 310024, P. R. China}

\author{Zhefeng Lou}
\affiliation{Key Laboratory for Quantum Materials of Zhejiang Province, Department of Physics, School of Science and Research Center for Industries of the Future, Westlake University, Hangzhou 310030, P. R. China} 
\affiliation{Institute of Natural Sciences, Westlake Institute for Advanced Study, Hangzhou 310024, P. R. China}

\author{Xiaohui Yang}
\affiliation{Key Laboratory for Quantum Materials of Zhejiang Province, Department of Physics, School of Science and Research Center for Industries of the Future, Westlake University, Hangzhou 310030, P. R. China} 
\affiliation{Institute of Natural Sciences, Westlake Institute for Advanced Study, Hangzhou 310024, P. R. China}
\affiliation{Department of Physics, China Jiliang University, Hangzhou 310018, Zhejiang, P. R. China}

\author{Zhiwei Wang}
\email{zhiweiwang@bit.edu.cn}
\affiliation{Centre for Quantum Physics, Key Laboratory of Advanced Optoelectronic Quantum Architecture and Measurement (MOE), School of Physics, Beijing Institute of Technology, Beijing 100081, China} 
\affiliation{Beijing Key Lab of Nanophotonics and Ultrafine Optoelectronic Systems, Beijing Institute of Technology, Beijing 100081, China}
\affiliation{Material Science Center, Yangtze Delta Region Academy of Beijing Institute of Technology, Jiaxing 314011, China}

\author{Yugui Yao}
\affiliation{Centre for Quantum Physics, Key Laboratory of Advanced Optoelectronic Quantum Architecture and Measurement (MOE), School of Physics, Beijing Institute of Technology, Beijing 100081, China} 
\affiliation{Beijing Key Lab of Nanophotonics and Ultrafine Optoelectronic Systems, Beijing Institute of Technology, Beijing 100081, China}
\affiliation{Material Science Center, Yangtze Delta Region Academy of Beijing Institute of Technology, Jiaxing 314011, China}

\author{Congjun Wu}
\email{wucongjun@westlake.edu.cn}
\affiliation{Key Laboratory for Quantum Materials of Zhejiang Province, Department of Physics, School of Science and Research Center for Industries of the Future, Westlake University, Hangzhou 310030, P. R. China} 
\affiliation{Institute of Natural Sciences, Westlake Institute for Advanced Study, Hangzhou 310024, P. R. China}
\affiliation{Institute for Theoretical Sciences, Westlake University, 310024, Hangzhou, China}
\affiliation{New Cornerstone Science Laboratory, Department of Physics, School of Science, Westlake University, 310024, Hangzhou, China}

\author{Xiao Lin}
\email{linxiao@westlake.edu.cn}
\affiliation{Key Laboratory for Quantum Materials of Zhejiang Province, Department of Physics, School of Science and Research Center for Industries of the Future, Westlake University, Hangzhou 310030, P. R. China} 
\affiliation{Institute of Natural Sciences, Westlake Institute for Advanced Study, Hangzhou 310024, P. R. China}

\date{\today}

\begin{abstract} 
\noindent
The interplay among frustrated lattice geometry, nontrivial band topology and correlation yields rich quantum states of matter in Kagome systems~\cite{Balents2010Nature,WenXG2009PRB}. A series of recent members in this family, $A$V$_3$Sb$_5$ ({$A$}= K, Rb, Cs), exhibit a cascade of symmetry-breaking transitions~\cite{Zeljkovic2021nature}, involving the 3Q chiral charge ordering~\cite{Miao2021PRX,Wilson2021PRX,Guguchia2022Nature,Hasan2021NM,Moll2022Nature}, electronic nematicity~\cite{ChenXH2022Nature,Zeljkovic2022NP}, roton pair-density-wave~\cite{GaoHJ2021Nature} and superconductivity~\cite{Wilson2020PRL}. The nature of the superconducting order is yet to be resolved. 
Here, we report an indication of chiral superconducting domains with boundary supercurrents in intrinsic CsV$_3$Sb$_5$ flakes. Magnetic field-free superconducting diode effect is observed with polarity modulated by thermal histories, suggesting dynamical superconducting order domains in a spontaneous time-reversal symmetry breaking background. 
Strikingly, the critical current exhibits the double-slit superconducting interference patterns when subjected to an external magnetic field. Characteristics of the patterns are modulated by thermal cycling. 
These phenomena are proposed as a consequence of periodically modulated supercurrents flowing along certain domain boundaries constrained by fluxoid quantization. Our results imply a chiral superconducting order, opening a potential for exploring exotic physics, e.g. Majorana zero modes, in this intriguing topological Kagome system.



\end{abstract}

\maketitle


\noindent
Chiral superconductors (SC), characterized by complex order parameters, break time-reversal symmetry (TRS) spontaneously. Certain types of chiral SCs are topologically non-trivial whose gap functions exhibit phase winding over the Fermi surface. They allow for exploring chiral edge states and Majorana zero modes, showing a promise for fault-tolerant topological quantum computation~\cite{Ivanov2001PRL}. A well-known example exhibiting chiral pairing symmetry is the superfluid $^3$He-A phase~\cite{Leggett1975RMP}. In electronic materials, evidence of TRS breaking has been reported~\cite{Kallin2016RPP,Kapitulnik2014Science,JiaoL2020Nature,Weitering2023NP}, including Sr$_2$RuO$_4$~\cite{Kallin2016RPP}, UPt$_3$~\cite{Kapitulnik2014Science} and UTe$_2$ ~\cite{JiaoL2020Nature}.
Nevertheless, an unequivocal demonstration of chiral edge supercurrent ($I_\textrm{e}$) at SC domain boundaries is still a pending target.

$A$V$_3$Sb$_5$ ($A$=K, Rb, Cs) exhibit a rich phase diagram featured by the intricate interplay among 
multiple intertwined orders \cite{Miao2021PRX,Wilson2021PRX,Guguchia2022Nature,Hasan2021NM,Moll2022Nature,ChenXH2022Nature,Zeljkovic2022NP,Zeljkovic2023NP,GaoHJ2021Nature}. This hints the possibility of unconventional SC~\cite{ChenXH2022Nature1,Hasan2021NM,GaoHJ2021Nature}, yet the pairing symmetry remains unclear~\cite{LiSY2021Arxiv,Khasanov2023NC,LuoJL2021CPL,Yuan2021SCPMA,Shibauchi2023NC,Okazaki2023Nature,FengDL2021PRL}. Accumulated evidence~\cite{LuoJL2021CPL,Yuan2021SCPMA,Shibauchi2023NC,Okazaki2023Nature,FengDL2021PRL}, including nuclear quadrupole resonance~\cite{LuoJL2021CPL}, tunnel diode oscillator~\cite{Yuan2021SCPMA}, electron irradiation~\cite{Shibauchi2023NC} and angle-resolved photoemission spectroscopy~\cite{Okazaki2023Nature}, surprisingly indicates the presence of nodeless, spin-singlet, and nearly isotropic SC gaps. From a theoretical perspective, a chiral SC state with a fully gapped composite gap function has been proposed in literature ~\cite{LiJX2012PRB,Thomale2021PRL,Anderson2022PRB}. Relevant clues to this are elusive~\cite{Guguchia2022Nature,Khasanov2022CP,Khasanov2023NC,Okazaki2023Nature}, derived exclusively from muon spin spectroscopy ($\mu$SR) that detects signals of broken TRS. Phase-sensitive probes are crucial but still lacking. 
Here, we present signatures of chiral SC domains in CsV$_3$Sb$_5$ via phase-sensitive measurements.


\begin{figure*}[thb]
	\begin{center}
		\includegraphics[width=18cm]{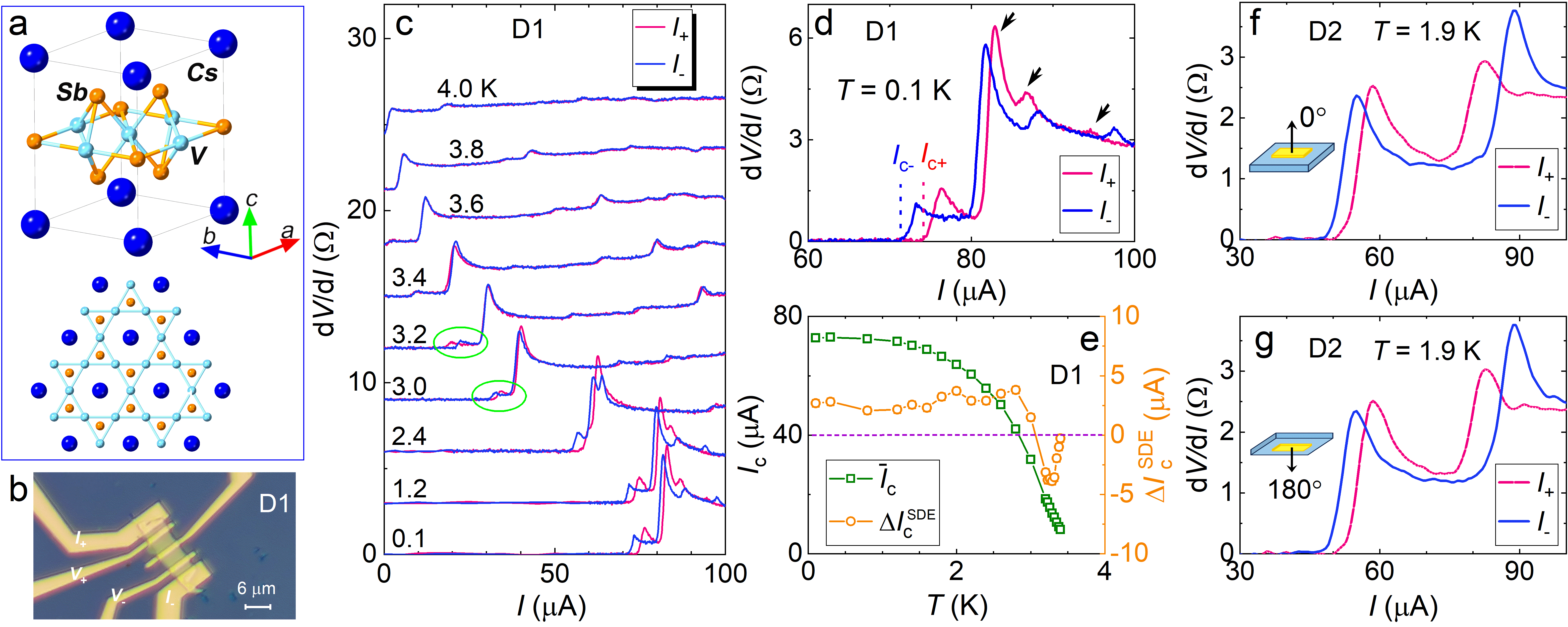}
	\end{center}
	\setlength{\abovecaptionskip}{-8 pt}
	\caption{\textbf{Zero-field superconducting diode.} \textbf{a}, Crystal structure of CsV$_3$Sb$_5$. \textbf{b}, Optical image of the device D1.  \textbf{c}, Differential resistance ($\textrm{d}V/\textrm{d}I$) as a function of d.c. current bias ($I$) at various $T$ for D1. 
    The red and blue curves are collected in positive ($I_\textrm{+}$) and negative ($I_\textrm{-}$) bias regimes, respectively. 
    Curves are offset from each other by 3 $\Omega$ for clarity. \textbf{d}, Enlarged curve of \textbf{c} at $T=0.1$~K. \textbf{e}, $T$-dependence of average critical current ($\bar{I_\textrm{c}}$) and $\Delta I^\textrm{SDE}_\textrm{c}$, where $\bar{I_\textrm{c}}=(I_\textrm{c+}+I_\textrm{c-})/2$ and $\Delta I^\textrm{SDE}_\textrm{c}=I_\textrm{c+}-I_\textrm{c-}$. \textbf{f} and \textbf{g}, $\textrm{d}V/\textrm{d}I$ versus $I$ for D2 with the setup at $0^\textrm{o}$ and $180^\textrm{o}$ configurations, respectively.     
	}
	\label{Fig1}
\end{figure*}

CsV$_3$Sb$_5$ hosts a hexagonal structure, composed of alternating stacks of V$_3$Sb$_5$ slabs and Cs layers, among which vanadium ions form the Kagome net, seen in Fig.~\ref{Fig1}a. For the measurements, three devices (D1-D3) were fabricated by mechanically 
exfoliating  nanoflakes from high-quality single crystals. The basic transport properties are presented in Extended Data Fig.~\ref{FigSRT}. 

~\\
\noindent\textbf{Zero-field superconducting diode effect}

\noindent
The chiral SC was inspected by observing the SC diode effect (SDE), which depicts an asymmetry of the critical current ($I_\textrm{c}$) with respect to the direction of current flow in the absence of TRS and inversion symmetry 
(IRS)~\cite{Ono2020Nature,WangXL2023NRP}. TRS and IRS are disrupted either internally or externally. 
In the case of internal TRS breaking, magnetic field ($B$)-free SDEs~\cite{Ali2022Nature,Parkin2022NM,Li2022NP} are realized, in turn reflecting the nature of pairing symmetry~\cite{WangXL2023NRP}. To examine intrinsic properties, non-SC contacts were made by gold deposition, seen in Fig.~\ref{Fig1}b. The cryostat was warmed up to room temperature ($T$) to fully release residual flux trapped in the SC magnet prior to experiments. In Supplementary Note~1, additional procedures were implemented to further eliminate any remaining field. 

In Fig.~\ref{Fig1}c, the differential resistance (d$V$/d$I$) for D1 was measured by sweeping the d.c. current ($I$) at zero $B$ and various $T$. $I_\textrm{c}$ evolves with the reduction of $T$ along with a noticeable inequivalence between the positive ($I_\textrm{+}$) and negative ($I_\textrm{-}$) bias regimes. A magnified curve at $T=0.1~$K is specified in Fig.~\ref{Fig1}d, in which $I_\textrm{c+}$  along the positive direction is larger than $I_\textrm{c-}$ along the negative one ($\Delta I^\textrm{SDE}_\textrm{c}=I_\textrm{c+}-I_\textrm{c-}\approx3~\mu$A), indicating non-reciprocity. Additionally, several non-reciprocal transition features (marked by arrows) are observed above $I_\textrm{c}$. 
Its relation to SC domain structures will be explained later. Fig.~\ref{Fig1}e displays the average $\bar{I_\textrm{c}}$ and $\Delta I^\textrm{SDE}_\textrm{c}$ versus $T$. As $T$ declines, $\Delta I^\textrm{SDE}_\textrm{c}$ shows a peculiar sign switching slightly below the SC transition temperature ($T_\textrm{c}\approx3.5$~K), as indicated by circles in Fig.~\ref{Fig1}c, and eventually its polarity becomes stable. This polarity variation is weird, distinct from what was reported~\cite{WangXL2023NRP,Parkin2022NM}. In Extended Data Fig.~\ref{FigSSDET}, the SDE polarity could be reversed after thermal cycling from $T$ slightly above $T_\textrm{c}$, which also shows the magnitude alteration of $\bar{I_\textrm{c}}$ and $\Delta I^\textrm{SDE}_\textrm{c}$. 
Relevant data, including  half-wave rectification, are shown in Extended Data Fig.~\ref{FigSHWR}, \ref{FigSIV}. 
It is noteworthy that the SDE signals remained unchanged even when the setup was reversed in Fig.~\ref{Fig1}f and g, underscoring the prominent role played by the spontaneous TRS-breaking, instead of from the environment.
See more discussion in Supplementary Note~1.


The presence of thermal switching and polarity variation in zero-field SDEs hints the existence 
of dynamic orders in an internal TRS-breaking background. 
A plausible candidate is chiral SC domains, 
which will be further elucidated below. 
This finding is hardly explained by the chiral charge order inherited from the normal state
(see discussion in Methods). 




~\\ \noindent\textbf{Superconducting interference patterns}

\begin{figure*}[thb]
	\begin{center}
		\includegraphics[width=18cm]{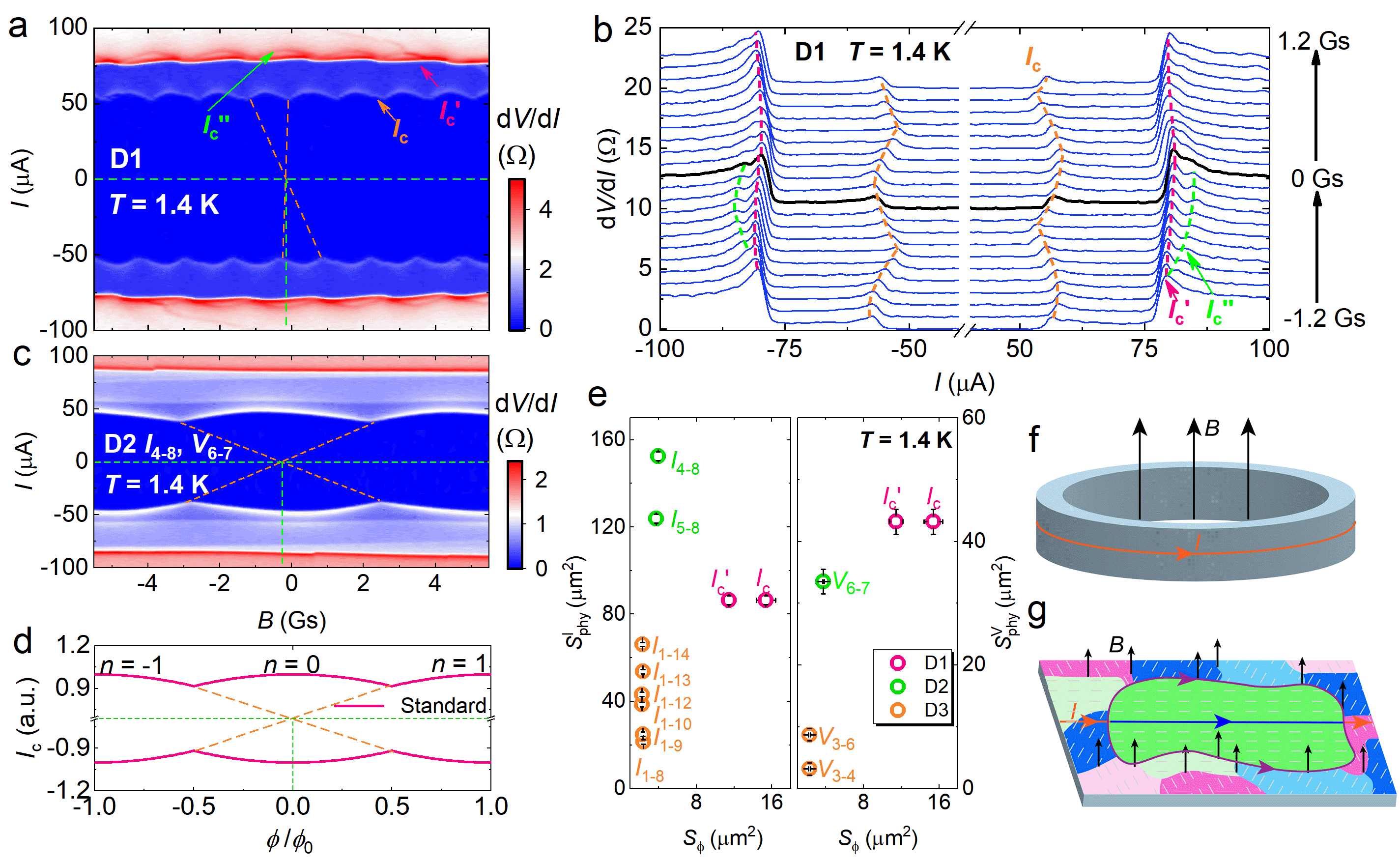}
	\end{center}
	\setlength{\abovecaptionskip}{-8 pt}
		\caption{\textbf{Superconducting interference patterns on intrinsic CsV$_3$Sb$_5$ flakes.} \textbf{a}, SIPs on D1 measured at 1.4 K. See the patterns at a broader scale in Extended Data Fig.~5. \textbf{b}, Corresponding sets of $\textrm{d}V/\textrm{d}I$ versus $I$ at $-1.2<B<1.2$~Gs. Three sets of SIPs in \textbf{a} are traced out by the transition anomalies, delineated by dashed curves and denoted as $I_\textrm{c}$, $I'_\textrm{c}$ and $I''_\textrm{c}$. \textbf{c}, SIPs on D2 measured at 1.4 K. \textbf{d}, Standard SIP derived from Eq.~\ref{Eq1}. The orange dashed lines in \textbf{a}, \textbf{c} and \textbf{d} connect the minimum of the oscillation profiles within the $n=0$ segment. Their intersection, offset from  $B=0$~Gs, reveals the deviation from the standard model. \textbf{e}, Physical area ($S^\mathrm{I/V}_\textrm{phy}$) versus the flux penetration area ($S_{\phi}$) for D1, D2 and D3. $S^\mathrm{I/V}_\textrm{phy}$ is the area between a couple of current/voltage electrodes denoted by a numerical pair (see Fig. \ref{Fig1}b and Extended Data Fig.~\ref{FigSIV},~\ref{FigSSIPD123}). The data was extracted from Fig.~\ref{Fig2}a, c and Extended Data Fig.~\ref{FigSSIPvI},~\ref{FigSSIPvV}. The error bars indicate uncertainties in the determination of $S^\mathrm{I/V}_\textrm{phy}$ and $S_{\phi}$. $S_{\phi}$ is insensitive to the variation of current or voltage electrodes, e.g. for D3, implying that $S_{\phi}$ is related to a specific area between two certain terminals. In Extended Data Fig.~\ref{FigSSIPvV}, the primary SIP for $V_\textrm{3-6}$ of D3 arises from regions between $V_\textrm{3-4}$, explaining the same $S_\mathrm{\phi}$ for $V_\textrm{3-6}$ and $V_\textrm{3-4}$ in panel right. \textbf{f}, Illustration of the LP device. \textbf{g}, Sketch of domain network in CsV$_3$Sb$_5$ devices. The supercurrent passes through the domains both along the boundaries and within the bulk. The pink, blue and green regions represent three types of $C_2$ rotational SC domains. Each hosts two degenerate phases of opposite chirality, as denoted by the heavy and light colors. Please refer to details in Discussion.
    }

	\label{Fig2}
\end{figure*}
\noindent 
In Fig.~\ref{Fig2}a, b, we measured $\textrm{d}V/\textrm{d}I$ at selective values of $B$ for D1 and plotted its color map in the $I$-$B$ plane. Intriguingly, three sets of periodic oscillation profiles ($I_\textrm{c}$, $I'_\textrm{c}$ and $I''_\textrm{c}$) are resolved in Fig.~\ref{Fig2}a, traced by three transition peaks marked by dashed curves in Fig.~\ref{Fig2}b. Similar patterns with distinct periodicity for D2 are presented in Fig.~\ref{Fig2}c. More data are
in Extended Data Fig.~\ref{FigSSIPD123}-\ref{FigSSIPvV}.
 


Such double-slit patterns vividly imitate the SC interference patterns (SIPs), which would be obtained from a Little-Parks (LP) device as depicted in Fig.~\ref{Fig2}d. In such a device, the magnetic flux threading in hollow regions modulates the loop supercurrent, as sketched in Fig.~\ref{Fig2}f~\cite{Barone1982book}. 
The fluxoid quantization within an enclosed area leads to the superfluid velocity $v_\textrm{e}=2\pi\hbar/(m^*L_c)(n-\phi/\phi_0)$, where $m^*$ is the Ginzburg-Landau (GL) effective mass; $L_c$ is the loop circumference; $n$ is the closest integer to $\phi/\phi_0$ with $\phi_0=hc/(2e)$ the flux quantum and $\phi=B S_{\phi}$ the flux enclosed by the loop;
$S_{\phi}$ is the field threading area. It subsequently leads to a modulation of the SC condensation wavefunction $\Psi_\textrm{e}$ as $\Delta |\Psi_\textrm{e}|^2\sim(n-\phi/\phi_0)^2$. Then, the variation of the GL critical current is yielded as (see more discussion in Supplementary Note 2):



\begin{equation}
\frac{\Delta I^\textrm{SIP}_\textrm{c}}{I_\textrm{c}} \sim -g\left(\frac{2\pi\xi}{L_\textrm{c}}\right)^2
\left(n-\frac{\phi}{\phi_0}\right)^2   
\label{Eq1}
\end{equation}
where $\xi$ is the GL coherence length and $g$ is a constant of order one.  

\begin{figure*}[thb]
\begin{center}
\includegraphics[width=18cm]{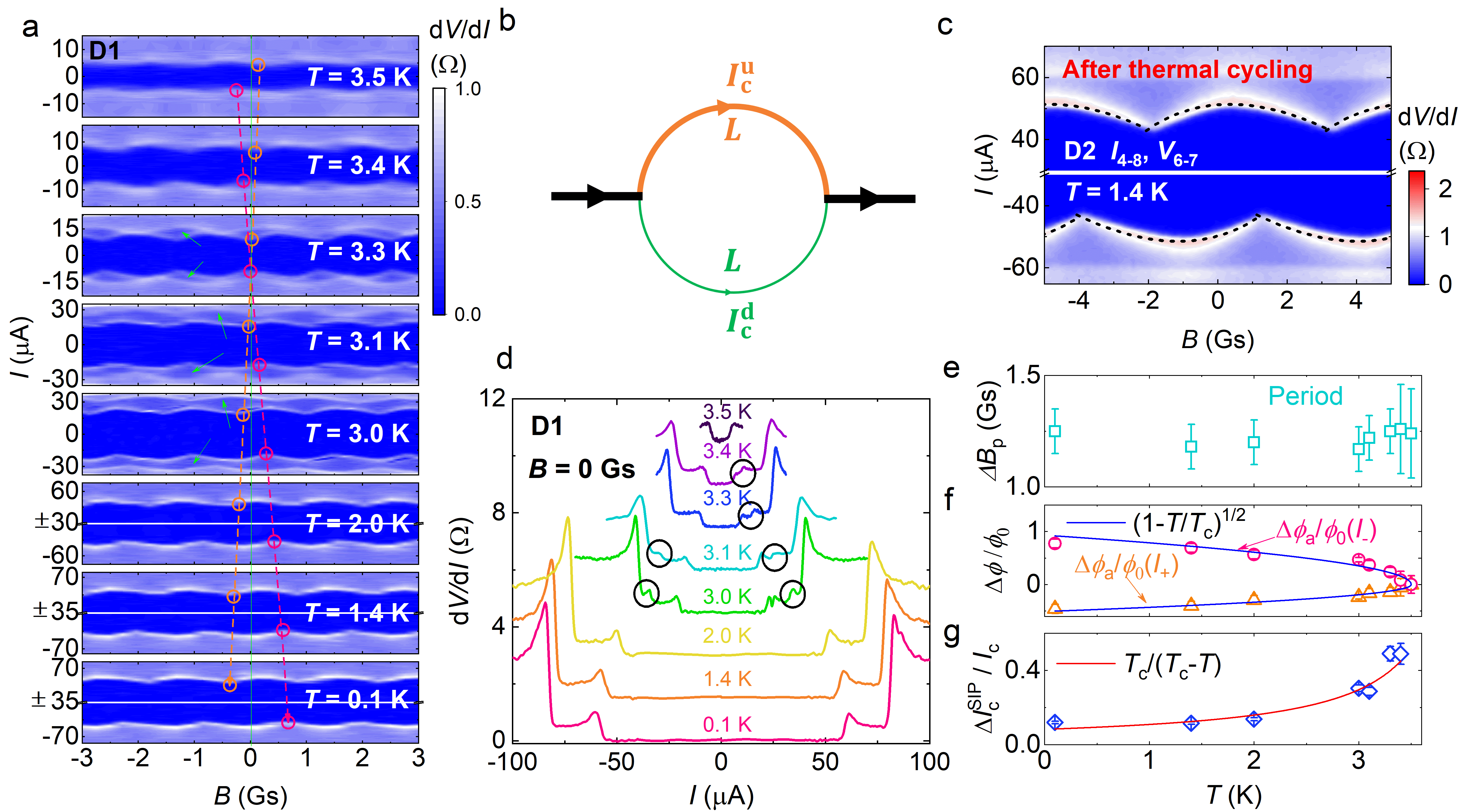}
\end{center}\setlength{\abovecaptionskip}{-8 pt}
\caption{
\textbf{Temperature evolution of SIPs for D1.} \textbf{a}, SIPs on D1 measured at various $T$. 
Open circles, spotted at the nearest minima of $I_\textrm{c}$ around 0 Gs,  tracks the counter-shift of $I_\textrm{+}$ and $I_\textrm{-}$ branches.  A new SIP appears at $T>3$~K marked by arrows.
\textbf{b}, Illustration of domain inversion asymmetry. The critical current passing through the upper and lower branches are unequal: $I^\textrm{u}_\textrm{c}\neq I^\textrm{d}_\textrm{c}$. $L$ is the inductance. 
\textbf{c}, Numerical simulation (dashed curves) of the observed SIPs in D2, incorporating domain inversion asymmetry. 
\textbf{d}, $\textrm{d}V/\textrm{d}I$ versus $I$ at $B=0$~Gs. A sudden peak emerges at $T>3$~K enclosed by circles, corresponding to the new SIP indicated in \textbf{a}.  
\textbf{e}, $T$-evolution of period ($\Delta B_\textrm{p}$). \textbf{f}, Relative phase counter-shift versus $T$. $\Delta \phi_\textrm{a}(I_\pm)$ is obtained by comparing $\phi_\textrm{a}(I_\pm)(T)$ with respect to the value at 3.5 K. The solid curves are fits to $\sqrt{1-T/T_\textrm{c}}$. \textbf{g}, $T$-dependence of normalized oscillation amplitude ($\Delta I^\textrm{SIP}_\textrm{c}/I_\textrm{c})$.  The solid curve is a fit to the GL theory. The error bars indicate uncertainties in determination of extracted values.
}
	\label{Fig3}
\end{figure*}

It seems counterintuitive that such phenomena exist in unpatterned SC devices as in our case. A plausible explanation may lie in the existence of edge supercurrent ($I_\textrm{e}$) forming closed loop structures at certain boundaries~\cite{Ong2020Science}. According to the oscillation period ($\Delta B_\textrm{p}$) extracted from Fig.~\ref{Fig2}a, c and Extended Data Fig.~\ref{FigSSIPvI},~\ref{FigSSIPvV}, $S_{\phi}$ is calculated through $S_{\phi}=\phi_0/\Delta B_\textrm{p}$ and is compared with $S^\mathrm{I/V}_\textrm{phy}$, the physical area enclosed by the current/voltage electrodes, in Fig.~\ref{Fig2}e. Note that $S_{\phi}$ is detected by varying the current (voltage) terminals, while keeping the voltage (current) terminals unchanged for a single device. $S_{\phi}$ is much smaller than $S^\mathrm{I}_\textrm{phy}$ and $S^\mathrm{V}_\textrm{phy}$. 
No scaling relationships are observed among them.  
This finding suggests that $I_\textrm{e}$ flows along a specific boundary within the sample~\cite{Ong2020Science}, aligning with the expectation of chiral SC domains. 


In a chiral SC, domain structures between opposite chiralities are often established due to spontaneous symmetry breaking~\cite{Sigrist1991RMP}. For CsV$_3$Sb$_5$, we will argue below that there is a network of domain walls guiding the flow of $I_\textrm{e}$ (see details in Discussion). 
At the walls, SC order parameters are relatively suppressed compared to the bulk~\cite{Weitering2023NP}. 
Thus, the magnetic flux passes through the walls and penetrates the domain on the scale of the Pearl length $\Lambda_\textrm{p}=2\lambda^2/d$, where $\lambda$ is the London length. Given $\lambda$(0~K)$\approx0.4~\mu$m ~\cite{Khasanov2022CP}, one yields $\Lambda_\textrm{p}(0~\textrm{K})\approx8$~$\mu$m for D1 with the thickness $d\approx40$~nm, which is longer than the domain length scale ($\sqrt{S_\phi}$), indicating the penetration of flux throughout the domain. The edge supercurrent flows along the wall such that the domain enclosed by the supercurrent loop acts like a hollow (weak shielding of $B$) that serves as a basis for the LP effect, as illustrated in Fig.~\ref{Fig2}g. {In Supplementary Note 3, we discuss that the circulating supercurrent occurring in the bulk does not induce oscillation patterns.} 


The preceding discussion primarily revolves around domain boundaries, by assuming a priori that the edge and bulk wave functions ($\Psi_\textrm{e}$ and $\Psi_\textrm{b}$) are treated separately. Indeed, the applied current ($I$) passes through both the domain bulk ($I_\textrm{b}$) and edge. As $I$ approaches to $I_\textrm{c}$, $I$ is self-distributed such that both the boundary and bulk regions tend toward the normal state. We propose that the sharp double-slit SIPs stems from the contribution of $I_\textrm{e}$, and $I_\textrm{b}$ gives rise to a slow, broad evolution of $I_\textrm{c}$, as shown in Extended Data Fig.~5b. In the figure, the double-slit SIP is enveloped by a broader Fraunhofer-like pattern, likely induced by a local Josephson junction between neighboring domains, displaying the superposition of domain edge and bulk contributions. See more discussion in the Supplementary Note~4, 5 and Methods.

\begin{figure*}[thb]
	\begin{center}
		\includegraphics[width=12cm]{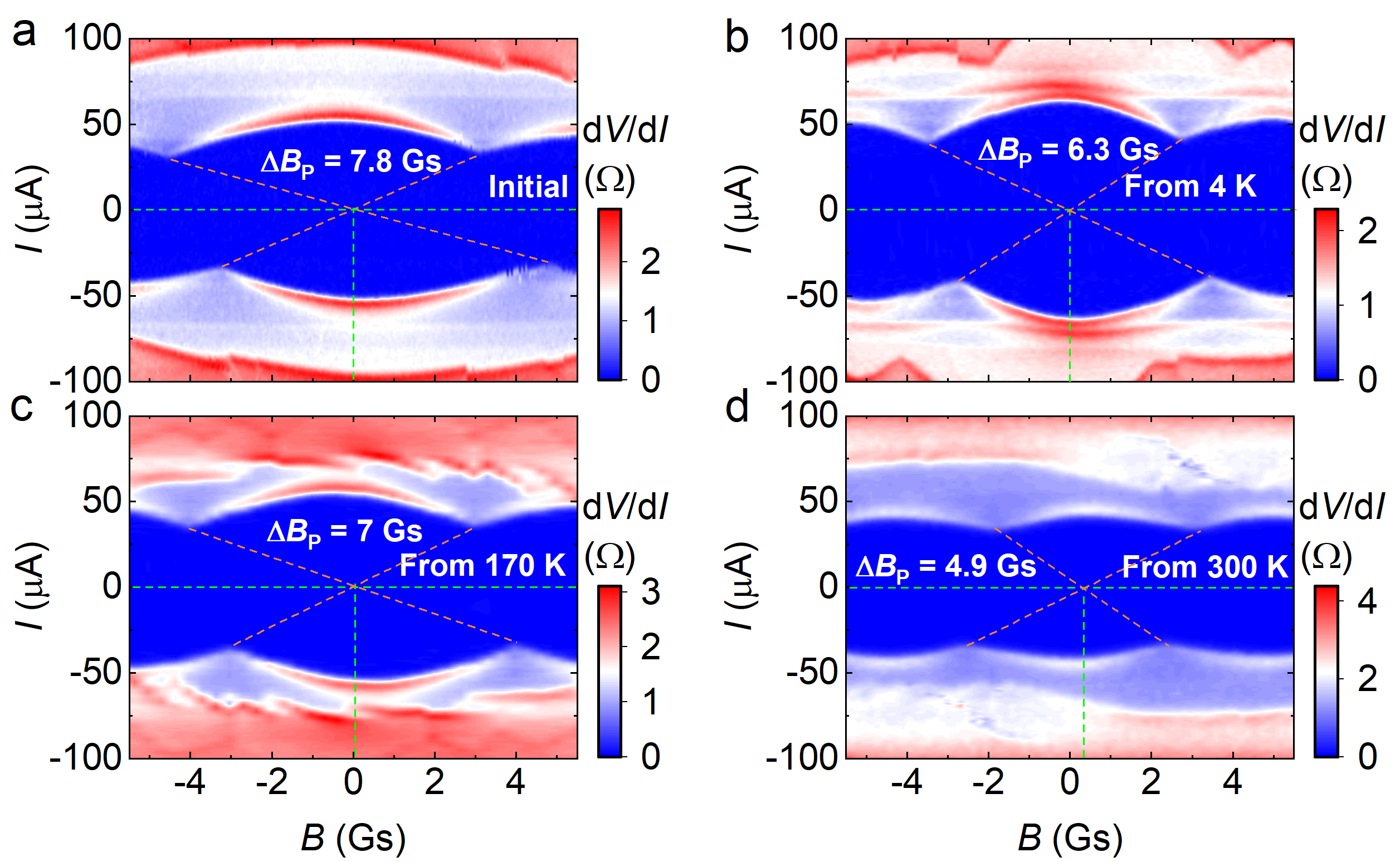}
	\end{center}
		\setlength{\abovecaptionskip}{-8 pt}
	\caption{\textbf{Thermal modulation of SIPs measured at 1.9~K for D2.} \textbf{a}, Initial SIPs. \textbf{b}, SIPs acquired from multiple thermal cycling from $T=4$~K, slightly above $T_\mathrm{c}$.  \textbf{c}, SIPs obtained from multiple thermal cycling from $T=170$~K, above $T_\textrm{CDW}$.  \textbf{d}, SIPs obtained from multiple thermal cycling from $T=300$~K. The dashed lines are guides to eyes. 
  }
	\label{Fig4}
\end{figure*}

Note that $I_\textrm{c}$ is not solely determined by the intrinsic critical current, but rather influenced by the connection strength between SC domains. The randomness in domain distribution and the spatial variation in the connection strength lead to the variation of $I_\textrm{c}$ across different voltage terminals in Extended Data Fig.~\ref{FigSIV} as well as the multiple transition features in Fig.~\ref{Fig1}d. The dynamics of domains account for the $I_\textrm{c}$ modulation after thermal cycling in Extended Data Fig.~\ref{FigSSDET}. In Extended Data Fig.~\ref{FigSSIPvV}, the sharp SIP exists only between terminals $V_\textrm{3-4}$, while others show obscure or even indiscernible patterns. This suggests specific domain structure requirements for observing SIPs, which, we consider, requires a SC domain of suitable dimensions (about 10~ $\mu$m$^{2}$) and regular geometries, preferably hosting a discrete $I_\textrm{c}$ distinct from the surrounding region in series. Discussion about the possibility of other scenarios is presented in Methods.
             

~\\
\noindent\textbf{Broken time-reversal and inversion symmetry}

\noindent 
Upon close inspection of SIPs in Fig.~\ref{Fig2}a, c, we discern a slight shift of the pattern symmetry center, 
marked by the intersection of dashed lines, compared to the standard model in Fig.~\ref{Fig2}d. 
Its deviation from zero field, defined as $B_\textrm{in}$, signifies a broken TRS~\cite{WuCJ2022Arxiv}. 
In Supplementary Fig.~3, SIPs at $0^\textrm{o}$ and $180^\textrm{o}$ configurations show indiscernible phase differences with the positive flux direction fixed relative to the flipped sample, indicating the predominant role of internal $B$ in determining the phase. In Supplementary Fig.~4, sweeping field within a range of $|B|<6$~Gs does not induce vortex trapping. Therefore, $B_\textrm{in}$ arises from an intrinsic TRS-breaking
effect.

The measured $B_\textrm{in}$ ranges from negligibly small to about 0.8~Gs in Supplementary Fig.~3a. The maximum is about threefold the internal magnetic field detected by $\mu$SR in the charge order phase~\cite{Guguchia2022Nature}, which is considered as a chiral flux phase composed of clockwise and anti-clockwise flux loops at the lattice sites in CsV$_3$Sb$_5$~\cite{HuJP2021SB}. $\mu$SR detects local field created by the imbalances of fluxes. The average $B_\textrm{in}$ at the micron scale in the normal state is thought to be largely eliminated, as supported by recent magneto-Kerr experiments that detected negligible TRS-breaking signals~\cite{XiaJ2023NC}. Moreover, the thermal modulation of the phase is shown below. Therefore, $B_\textrm{in}$ detected here stems, at least partially, from the SC order.  

Upon further scrutiny of Fig.~\ref{Fig2}a,c and~\ref{Fig3}a, we find the periodic profile is asymmetric within each segment of $I_{c+}$ and $I_{c-}$, i.e., lacking of a reflection symmetry. It also shows that $I_\text{c+}(B)\neq I_\text{c-}(B)$, indicating broken IRS~\cite{WuCJ2022Arxiv}. These features could be explained by asymmetry embedded in domain formation. 
As illustrated by the following model: The current loop is divided into two segments (up/down) with unequal critical current ($I^\textrm{u}_\textrm{c}\neq I^\textrm{d}_\textrm{c}$) in Fig.~\ref{Fig3}b due to the domain asymmetry.
The additional flux $\phi_\text{a}$ appears from the
self-inductance due to the imbalance between $I^\textrm{u}_\textrm{c}$
and $I^\textrm{d}_\textrm{c}$.

Obviously, $\phi_\text{a}$ is odd with respect to the current direction. Adding $\phi_\text{a}$ to Eq.~\ref{Eq1} produces an asymmetry in SIPs as shown by dashed lines in Fig.~\ref{Fig3}c that closely mimics the experimental results. 
More discussion of the asymmetric SIPs is presented in Supplementary Note~4. The presence of broken TRS and IRS in SIPs yields zero-field SDEs. 


Let us now gain more insights into the SIPs by analyzing the $T$-evolution. First, a new SIP emerges near $T_\textrm{c}$, marked by arrows in Fig. \ref{Fig3}a. It coincides with a sudden peak circled in the $\textrm{d}V/\textrm{d}I$ curves in Fig. \ref{Fig3}d, implying the formation of a new supercurrent loop. This highlights domain dynamics near $T_\textrm{c}$. 
Second, the period $\Delta B_\textrm{p}$($T$) is nearly constant within the experimental margin in Fig~\ref{Fig3}e, in contrast to the fact that $\Lambda_\textrm{p}(T)$ diverges as $T$ approaches $T_\textrm{c}$. This supports our argument that $S_\phi$ is truncated by the size of domains.
Third, the $I_\pm$ branches of a SIP exhibit an increasingly pronounced counter-shift in phase as $T$ lowers in Fig.~\ref{Fig3}a, accounting for the SDE evolution in Fig.~1e. In Fig.~\ref{Fig3}f, the counter-shift scales as  $\phi_\textrm{a}(I_\pm) 
\sim \mp\sqrt{1-T/T_\textrm{c}}$ as referred from Supplementary Note~4.
Finally, combining Eq.~\ref{Eq1} and the GL theory, one deduce the oscillation amplitude $\Delta I^\textrm{SIP}_\textrm{c}/I_\textrm{c}$$\sim\xi^2\sim T_\textrm{c}/(T_\textrm{c}-T)$,
which fits our results well in Fig.~\ref{Fig3}g. 

~\\
\noindent\textbf{Modulation of dynamic SC domains}

\noindent
The SC domain structure is closely linked with the distribution of local defects/strains and domain dynamics influenced by thermal histories. Accordingly, we observed the SIPs are not always identical after thermal cycling, but sometimes exhibit quantitative, or even qualitative, variations. Fig.~\ref{Fig4} shows remarkable thermal modulation of the domain size. $\Delta B_\mathrm{p}$ evolves from an initial value of 7.8~Gs to 6.3~Gs after recooling the device from $4$~K, slightly above $T_\mathrm{c}$. Subsequent cooling from $170~\mathrm{K}$ returns $\Delta B_\mathrm{p}$ to 7~Gs. Further cooling from 300~K significantly alters $\Delta B_\mathrm{p}$ to 4.9~Gs. Moreover, the patterns beyond the primary profile are substantially altered after thermal cycling. Combining Fig.~\ref{Fig4} and Supplementary Fig.~3, 5, we also observe thermal modulation of phase and domain asymmetry. 
All these provide remarkable indications of the existence of dynamic SC domains. In Fig.~\ref{Fig4}b, 
$\Delta{B_\mathrm{p}}$ shows noticeable changes although the thermal cycling is well below
the charge ordering temperature ($T_\textrm{CDW}\approx80$~K). Hence, the influence of charge order on the distribution of SC domains is not obvious.
Moreover, the field modulation of SIPs is presented in Extended Data Fig.~\ref{EDF9} and \ref{EDF10}.
They show either enhanced $|I_\mathrm{c}|$ or ``advanced" nature of hysteresis during field sweeps after field cooling, which is likely related to the rearrangement of SC domain walls driven by external $B$~\cite{Harlingen2006Science}.




~\\
\noindent\textbf{Discussion}

\noindent 
The cumulative evidence from SDEs and SIPs supports the existence of a TRS breaking SC order in CsV$_3$Sb$_5$. 
Nevertheless, the specific nature of its pairing state remains an open question. 
As discussed in Supplementary Note~6, previous theoretical studies have proposed a chiral $d_{\mathrm{x}^2-\mathrm{y}^2}\pm i d_{\mathrm{xy}}$ pairing symmetry 
characterized by a full SC gap~\cite{LiJX2012PRB,Thomale2021PRL,Anderson2022PRB}, which could reconcile our results with the majority of prior experimental outcomes~\cite{LuoJL2021CPL,Yuan2021SCPMA,Shibauchi2023NC,Okazaki2023Nature,FengDL2021PRL}. 
However, the double-degenerate orders in a $d\pm id$ SC are insufficient to form a network of chiral domains. 

In the normal state of CsV$_3$Sb$_5$, the nematicity already disrupts the $C_6$ rotation symmerty of the electronic structure, transforming it into $C_2$~\cite{Zeljkovic2023NP}. There is possibility that SC developed from this state inherently possesses a nematic ground state. In this case, three types of $C_2$ SC domains, each accommodating two degenerate phases of opposite chirality, constitute sixfold domains, potentially responsible for the network configuration illustrated in Fig.~\ref{Fig2}g. The supercurrents flowing along  domain boundaries lead to the observation of dynamic SIPs. Further details are given in Supplementary Note~7.


The method employed in this study may offer a simple probe for detecting domain edge supercurrents, contributing to the exploration of chiral SCs. The micro-sized SC domains could be further examined by high-resolution scanning probes capable of imaging current distributions. In the end, we note other scenarios, including pair-density-wave phase~\cite{GaoHJ2021Nature}, chiral charge order~\cite{Guguchia2022Nature} or significant inhomogeneity, may explain certain aspects of our findings. Their possibilities are evaluated by extended discussion in Methods. A complete understanding of all these issues still requires further efforts.

~\\
\noindent\textbf{Online content}

\noindent Supplementary materials are available at the online version of the paper.

~\\



~\\
\small
\noindent\textbf{Acknowledgments}
The authors are grateful to Lin Jiao and Chunyu Guo for the helpful discussion.  
This research is supported by Zhejiang Provincial Natural Science Foundation of China for Distinguished Young Scholars under Grant No. LR23A040001. C.W. is supported by the National Natural Science Foundation of China under the Grants No. 12234016 and No. 12174317. T.L. acknowledges support from the China Postdoctoral Science Foundation (Grant No. 2022M722845 and No. 2023T160586). This work has been supported by the New Cornerstone Science Foundation. X.L. acknowledges the support by the Research Center for Industries of the Future (RCIF) at Westlake University under Award No. WU2023C009. The authors thank the support provided by Dr. Chao Zhang from Instrumentation and Service Center for Physical Sciences at Westlake University.

~\\
\noindent\textbf{Author contributions}
T.L. fabricated the devices and did the transport measurements assisted by Z.X., J.W., Z.L. and X.Y.. J.L. prepared the samples supervised by Z.W. and Y.Y.. Z.P. did theoretical calculations supervised by C.W.. T.L., Z.P. and X.L. prepared the figures. X.L. wrote the paper with the inputs from T.L., Z.P. and C.W.. X.L. led the project. All authors contributed to the discussion.

~\\
\noindent\textbf{Competing interests}
The authors declare no competing interests.

~\\
\noindent\textbf{Additional information}

\noindent\textbf{Supplementary information} are available at the online version of the paper. 

\noindent Correspondence and requests for materials should be addressed to Zhiwei Wang, Congjun Wu or Xiao Lin.

\begin{appendix}
\setcounter{figure}{0}

\begin{figure*}[thb]
	\renewcommand{\figurename}{Extended Data Fig.}
	\includegraphics[width=16cm]{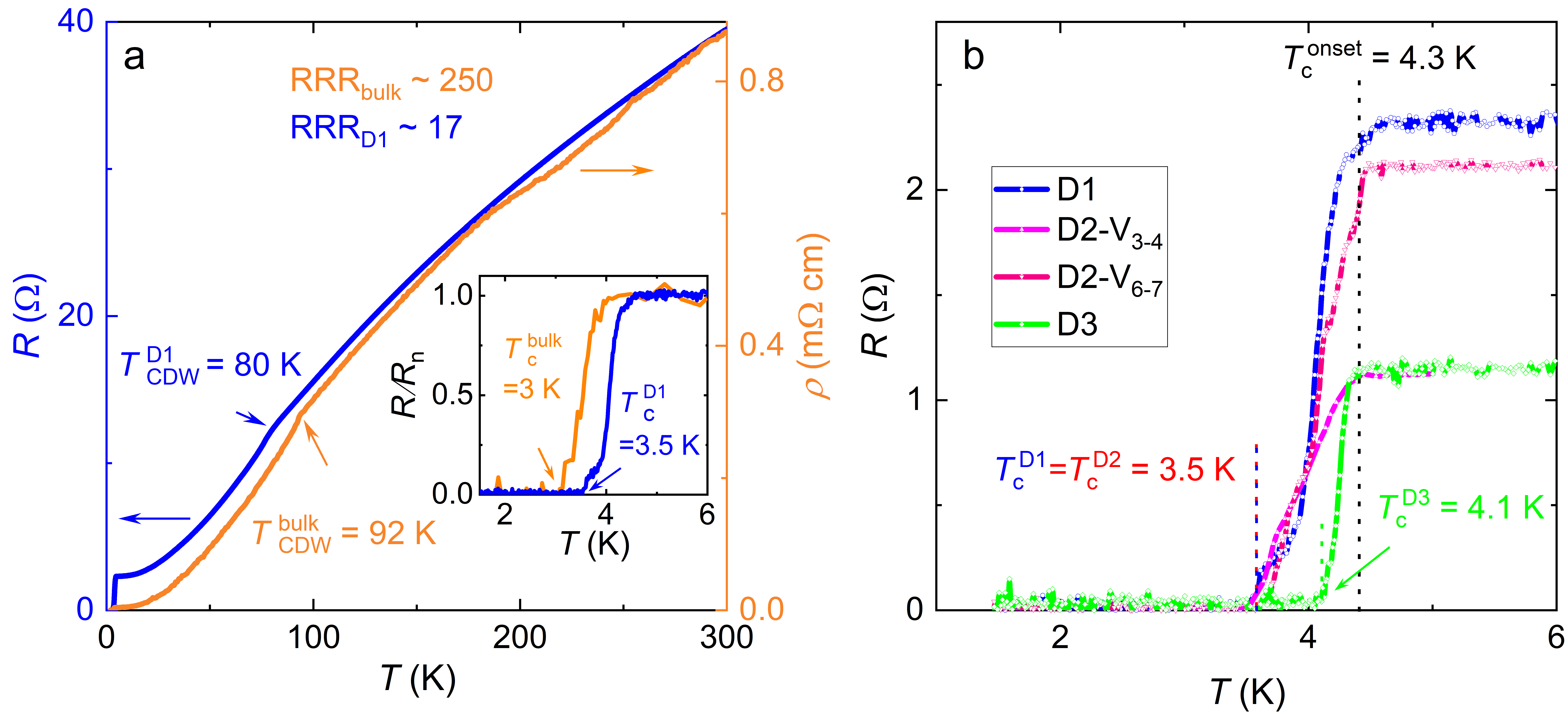}
	\caption{\textbf{Temperature dependence of resistance for D1-D3.} \textbf{a}, $T$-dependence of $\rho$ for the bulk single crystal (B1) and $R$ for a mechanically exfoliated specimen (D1) in the full-$T$ range.  The residual-resistance-ratio (RRR) amounts to 250 in B1, among the highest value of the literature~\cite{Wilson2021PRX,Wilson2020PRL,ChenXH2021PRX,LeiHC2021PRL}, highlighting the ultra-high quality of our crystals. The transition to chiral charge density wave (CDW) phase in B1 appears at $T_\textrm{CDW}\approx92$~K~\cite{Wilson2020PRL}, accompanied by a SC phase transition at $T_\textrm{c}\approx3$~K (determined at zero-resistance), consistent with previous reports~\cite{Wilson2020PRL}. In D1, $T_\textrm{CDW}$ is reduced to 80~K with the enhancement of $T_\textrm{c}$ to 3.5~K, as reported in the literature~\cite{LiSY2023NC}.      The inset presents the normalized resistance $R/R_\textrm{n}$ around $T_\textrm{c}$, where $R_\textrm{n}$ is the normal state resistance. \textbf{b}, $T$-dependence of $R$ around $T_\textrm{c}$ for D1-D3. For D2, we present the data collected at two sets of terminals: $V_\textrm{3-4}$ and $V_\textrm{6-7}$. The onset temperature of the superconducting transition ($T_\textrm{c}^\textrm{onset}$) for D1-D3 is similar, about 4.3~K. $T_\textrm{c}$ for D3 amounts to 4.1~K, which is higher than that of D1 and D2 (about 3.5~K). Note that $T_\textrm{c}$ of D2 measured at $V_\textrm{3-4}$ and $V_\textrm{6-7}$ is slightly different, reflecting different domain characteristics in-between the terminals. The thickness ($d$) of D1-D3 is about 40 nm.}
	\label{FigSRT}
\end{figure*}

\begin{figure*}[thb]
	\renewcommand{\figurename}{Extended Data Fig.}
	\includegraphics[width=18cm]{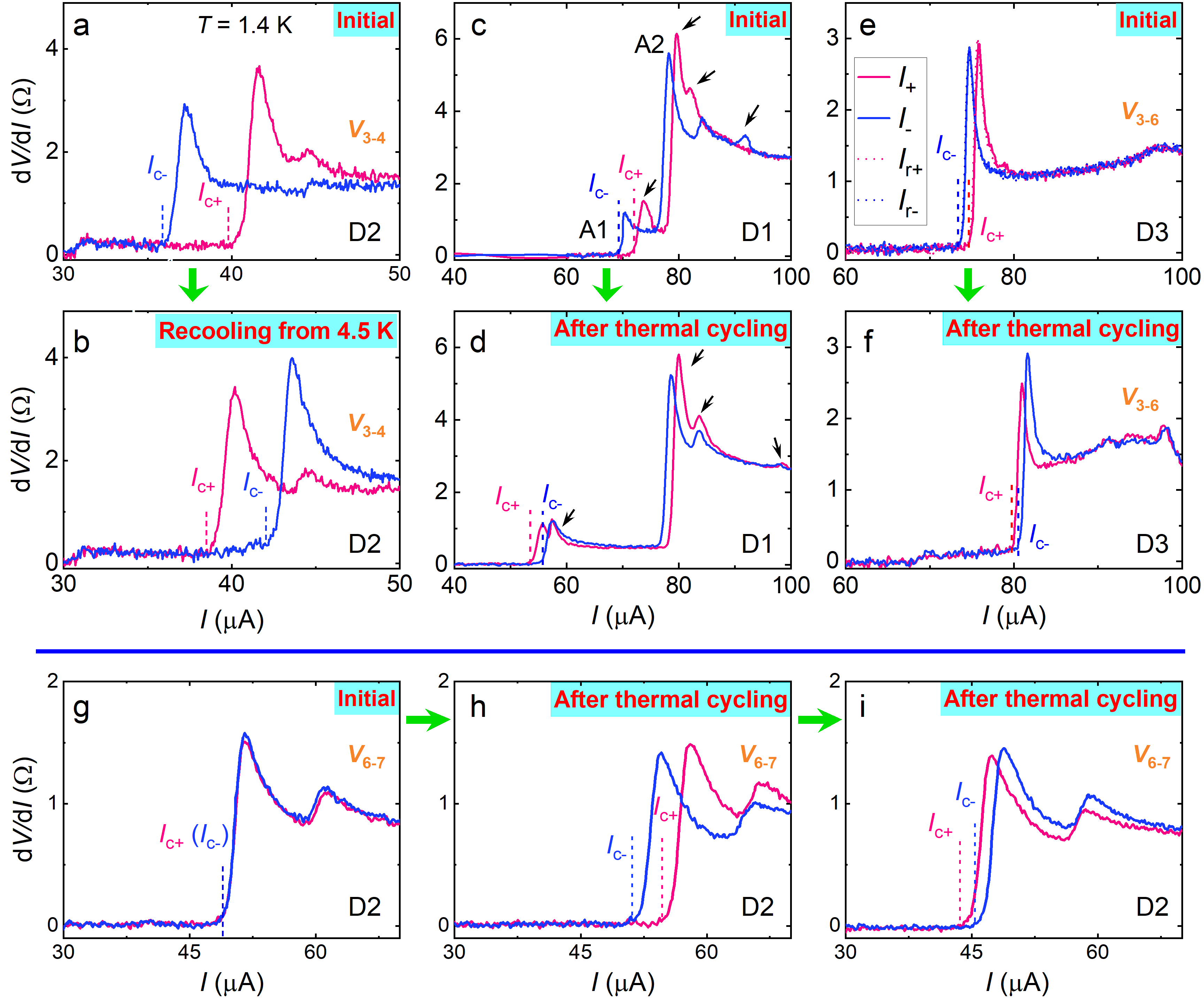}
    \caption{\textbf{Thermal modulation of zero field SDE for D1-D3 measured at $T=1.4$~K.} \textbf{a-b}, $\textrm{d}V/\textrm{d}I$ versus $I$ for terminals $V_\textrm{3-4}$ of D2 before (\textbf{a}) and after recooling from 4.5~K, slightly above $T_\textrm{c}$ (\textbf{b}). \textbf{c-d}, $\textrm{d}V/\textrm{d}I$ versus $I$ for D1 before (\textbf{c}) and after thermal cycling (\textbf{d}). \textbf{e-f}, $\textrm{d}V/\textrm{d}I$ versus $I$ for terminal $V_\textrm{3-6}$ of D3 before (\textbf{e}) and after thermal cycling (\textbf{f}).   In \textbf{e}, the measurement includes four branches: sweeping $I$ from zero to positive ($I_\textrm{+}$), from positive back to zero ($I_\textrm{r+}$), from zero to negative ($I_\textrm{-}$) and from negative back to zero ($I_\textrm{r-}$). The hysteresis between  $I_\textrm{+}$ ($I_\textrm{-}$) and $I_\textrm{r+}$ ($I_\textrm{r-}$) is negligible, indicating the absence of thermal heating or current re-trapping effect. Below, several observations are made: 1. All the devices exhibit remarkable non-reciprocity. 2. Not only the polarity, but also the magnitude of $\Delta I^\textrm{SDE}_\textrm{c}$ and $\bar{I_\textrm{c}}$ could be changed by thermal cycling.  3. The curves in \textbf{c} and \textbf{d} show multiple transition-like features with  non-reciprocity (marked by arrows), probably related to the difference in $I_\textrm{c}$ across different superconducting domain boundaries. 4. In \textbf{c} and \textbf{d}, the SDE polarity at A1 is reversed after thermal cycling, while the polarity at A2 remains unchanged. As discussed in Methods, the dynamic nature of SDEs with multiple transition peaks is unlikely to be fully explained by scenarios involving the combination of chiral charge order and certain sources of IR breaking such as geometric asymmetry and significant sample inhomogeneity. While, all of these could be reconciled with the existence of dynamic SC domains with broken TRS. Characteristics of the domains, such as domain asymmetry and inter-domain interaction, are randomly altered by thermal cycling (i.e. recooling the system from above $T_\textrm{c}$). \textbf{g-i} $\textrm{d}V/\textrm{d}I$ versus $I$ for terminals $V_\textrm{6-7}$ of D2 before (\textbf{g}) and after  thermal cycling (\textbf{h,i}). $V_{6-7}$ shows negligible non-reciprocity in the initial state. After thermal cycling, a finite SDE with either positive and negative polarity is induced.}
	\label{FigSSDET}
\end{figure*}

\begin{figure*}[thb]
	\renewcommand{\figurename}{Extended Data Fig.}
	\includegraphics[width=14cm]{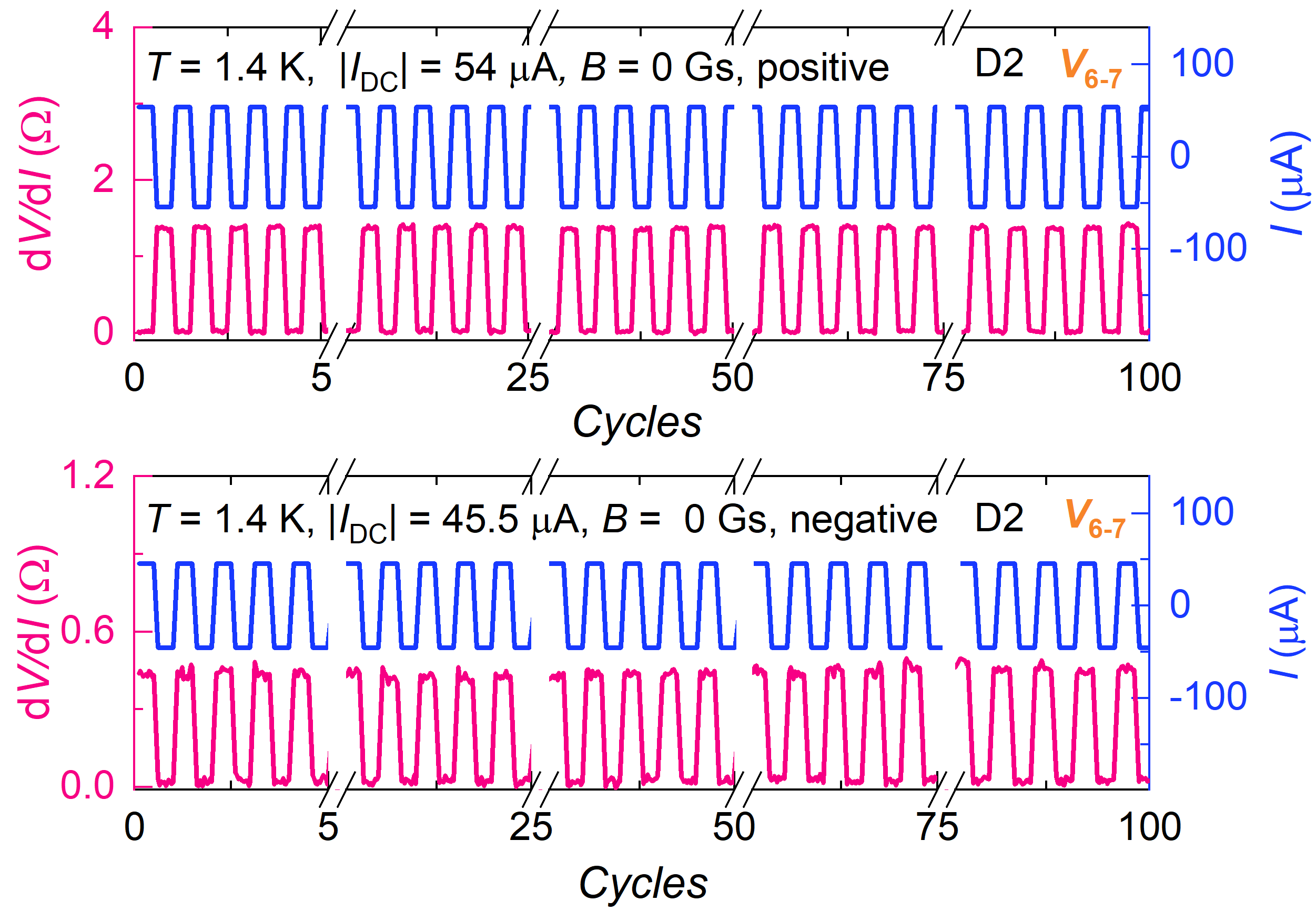}
	\caption{\textbf{Demonstration of half-wave rectification.} Direction-selective supercurrent transmission is demonstrated at $V_{6-7}$ of D2 with positive (upper panel) and negative polarity (lower panel). The measurements were performed by alternating the current polarity every 15.5 seconds. SDE remains stable after 100 cycles.
    }		
		\label{FigSHWR}
	\end{figure*}

\begin{figure*}[thb]
	\renewcommand{\figurename}{Extended Data Fig.}
	\includegraphics[width=10cm]{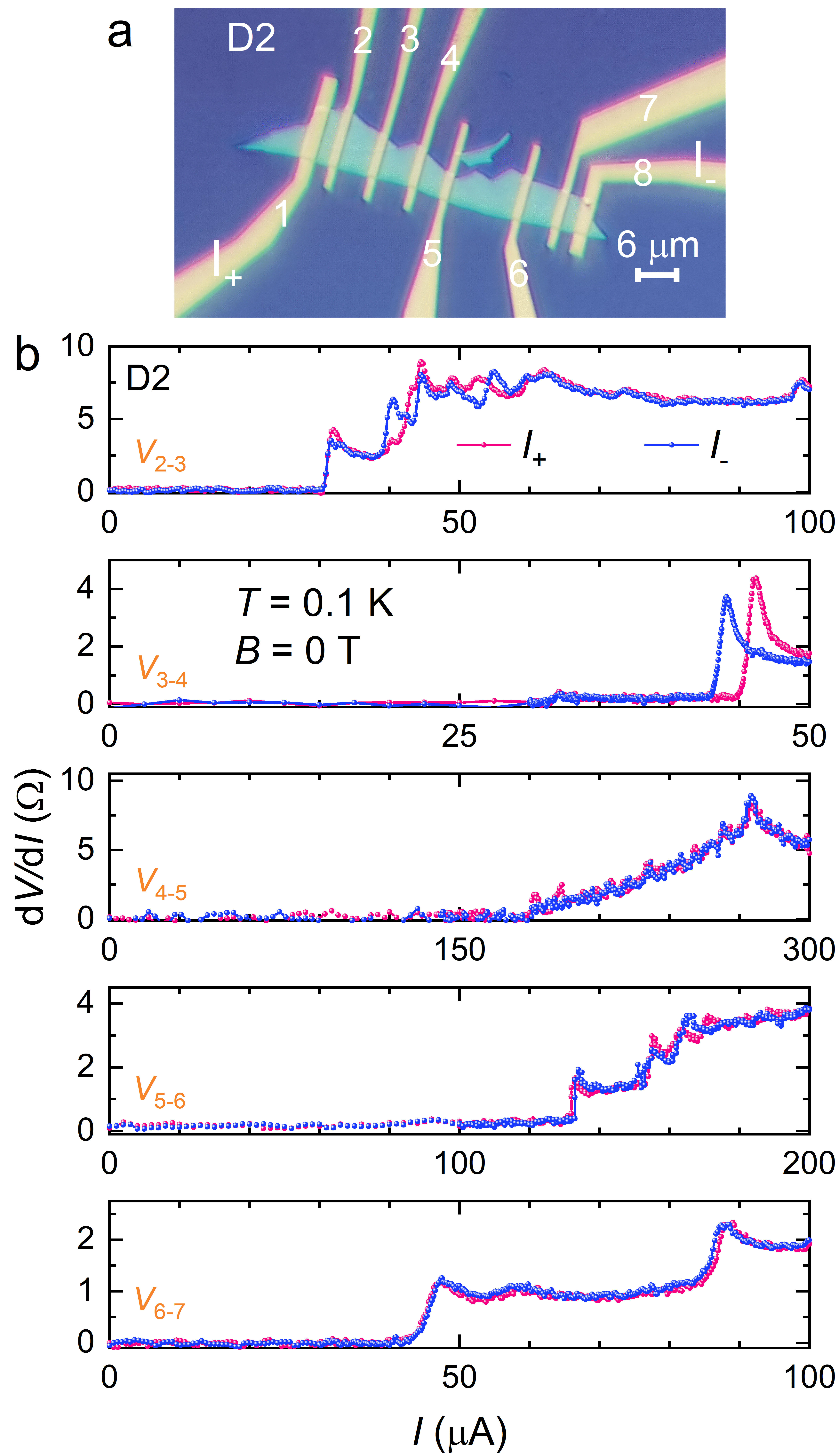}
	\caption{\textbf{Multi-step SC phase transitions on d$V$/d$I$ versus $I$ for D2 measured at different terminals}. \textbf{a}, Optical image of D2.  All the terminals are numbered (1-8), where 1 and 8 are for current and others are for voltage. \textbf{b}, d$V$/d$I$ versus $I$ at various terminals. Characteristics of d$V$/d$I$ exhibit notable distinctions across different terminals, including the variation of  $I_\textrm{c}$. Only $V_{3-4}$ displays an apparent SDE. Note that the SDE at $V_{6-7}$ can be excited by thermal cycling, as shown in Extended Data Fig.~2\textbf{g-i}. Given the ultra-high quality of CsV$_3$Sb$_5$ single crystals, mild device fabrication processes and the dynamic features on d$V$/d$I$, the multi-step transitions and the variation of $I_\textrm{c}$ cannot be simply attributed to significant sample inhomogeneity, but rather implies the formation of SC domain structure (See more discussion in Methods). $I_\textrm{c}$ is influenced by the strength of inter-domain connections. 
	}	
	\label{FigSIV}
\end{figure*}
~\\

\begin{figure*}[thb]
	\renewcommand{\figurename}{Extended Data Fig.}
	\includegraphics[width=18cm]{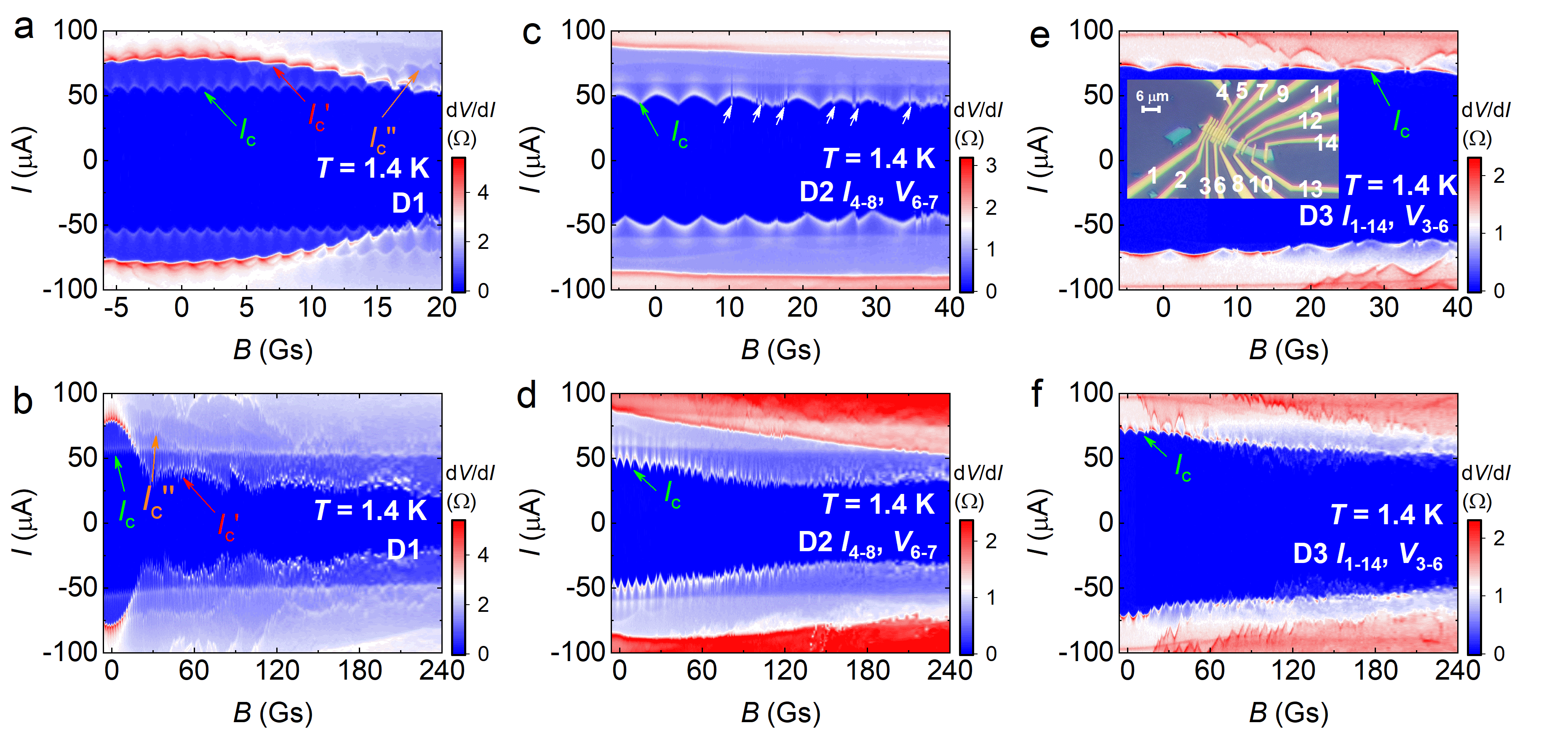}
	    \caption{\textbf{SIPs for D1, D2 and D3 in a broader range of $B$.} \textbf{a and b}, SIPs for D1, covering the $B$ range of 20~Gs and 240~Gs, respectively, as the Extended Data of Fig.~2a. In \textbf{a}, three SIPs ($I_\textrm{c}$, $I'_\textrm{c}$ and $I''_\textrm{c}$) are clearly resolved, corresponding to those in Fig.~2a. \textbf{b} displays more complex, periodic-like structures, alongside $I_\textrm{c}$, $I'_\textrm{c}$ and $I''_\textrm{c}$. Notably, we observe periodic oscillations on $I_\textrm{c}$. The magnitude of $I_\textrm{c}$ remains nearly unchanged in $B$ up to 240~Gs, as expected from the LP effect. In contrast, $I'_\textrm{c}$ displays a broad Fraunhofer-like pattern, on top of which is a rapid double-slit periodic oscillation. The broad feature is likely associated with a local Josephson junction between neighboring domains (domain bulk contribution) and the rapid mode arises from the LP effect from the domain edge. It implies a composite contribution from the bulk Josephson supercurrent and the edge supercurrent. Relevant discussion is also presented in the main text and Supplementary Note~4, 5. As discussed in Methods, such sharp double-slit SIPs are difficult to explain by alternative interpretations involving chiral charge orders or significant inhomogeneity with the absence of edge supercurrents. \textbf{c and d}, SIPs for D2, covering the $B$ range of 40~Gs and 240~Gs, respectively. An explicit periodic oscillation appears on $I_\textrm{c}$, along with some vague patterns. At $B>10$~Gs, distinct spikes (marked by white arrows) emerge on $I_\textrm{c}$, disrupting the periodic patterns, which is the result of the penetration of magnetic vortices into the domain bulk. \textbf{e and f}, SIPs for D3, covering the $B$ range of 40~Gs and 240~Gs, respectively. The inset of \textbf{e} is the optical image of D3. All the terminals are numbered (1-14).
     } 	
	\label{FigSSIPD123}
\end{figure*}

\begin{figure*}[thb]
	\renewcommand{\figurename}{Extended Data Fig.}
	\includegraphics[width=18cm]{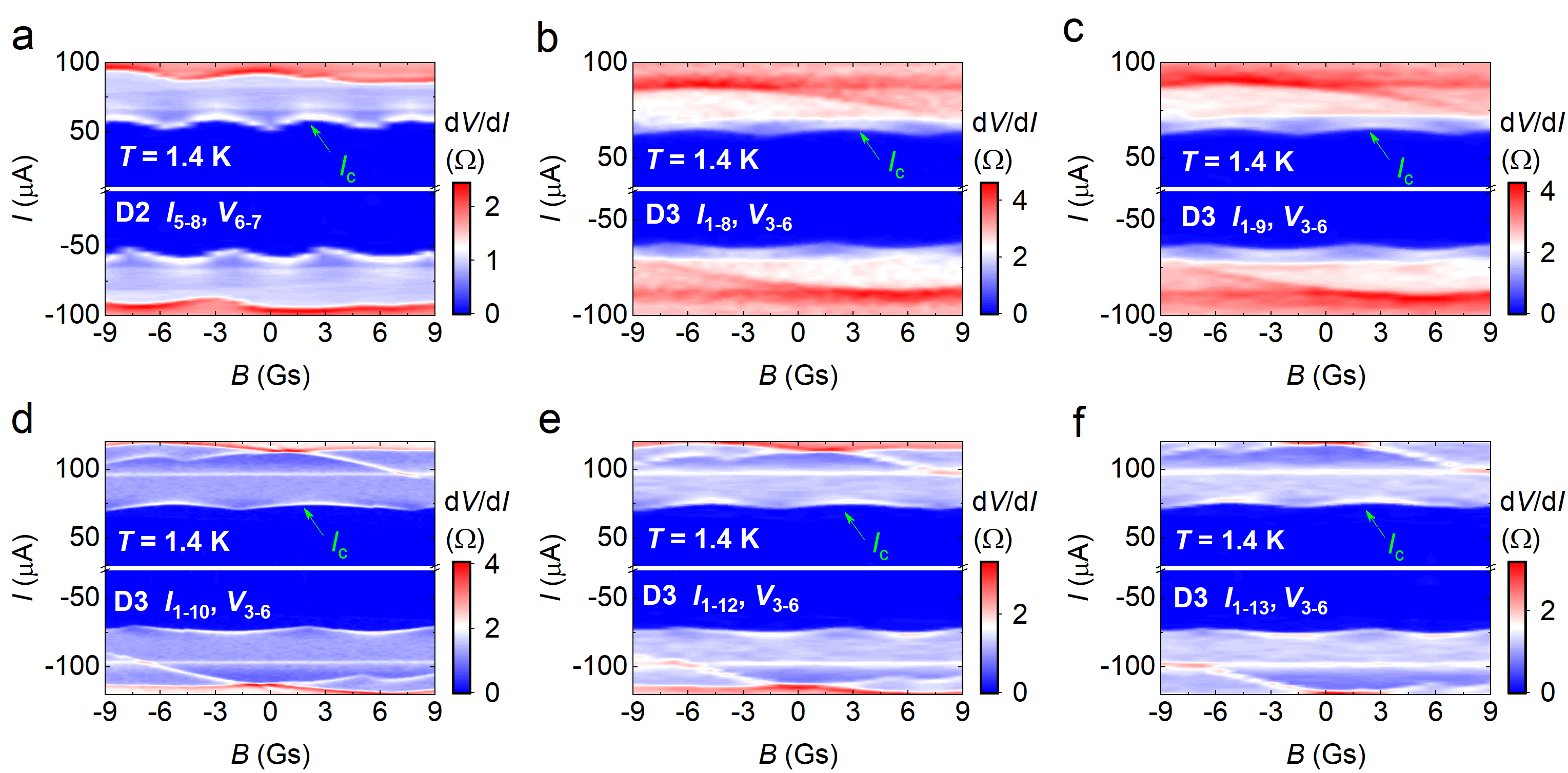}
	\caption{\textbf{SIPs for D2 and D3 with the current bias applied between different terminals.} \textbf{a}, SIPs for D2 measured at $V_\textrm{6-7}$ with current injected into $I_\textrm{5-8}$, which is compared with the data in Fig.~\ref{Fig2}c (D2, $V_\textrm{6-7}$ and $I_\textrm{4-8}$). \textbf{b-f}, SIPs for D3 measured at $V_\textrm{3-6}$, but with different current terminals. The oscillation patterns on $I_\textrm{c}$ is nearly unchanged when the current terminals are varied, indicating that the SIPs are associated with the domain structure between the voltage electrodes. 
	}	
	\label{FigSSIPvI} 
\end{figure*}

\begin{figure*}[thb]
	\renewcommand{\figurename}{Extended Data Fig.}
	\includegraphics[width=18cm]{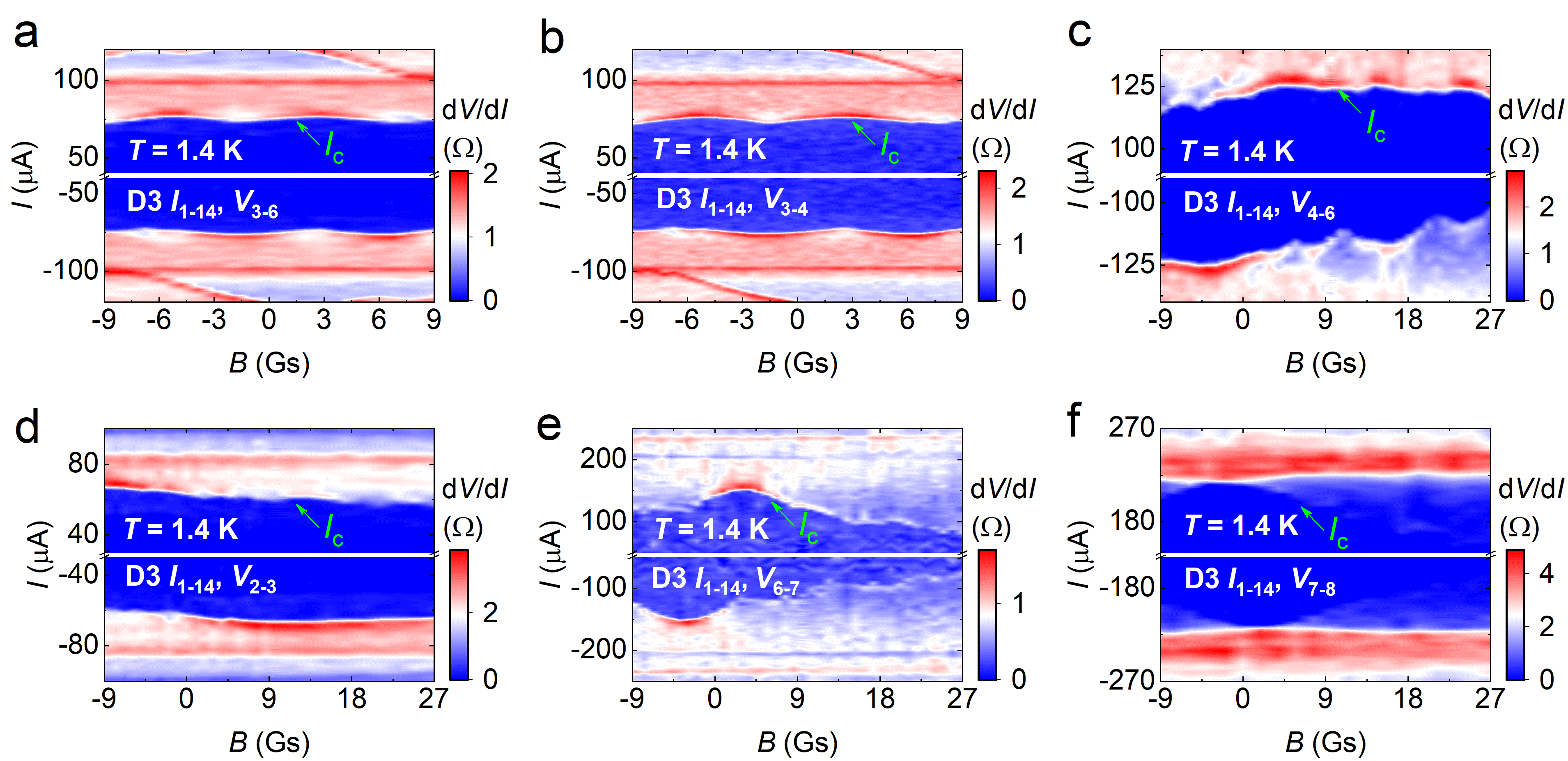}
		\caption{\textbf{SIPs for D3 collected at different voltage terminals.} \textbf{a-f}, SIPs measured by varying the voltage terminals while applying the current bias on $I_\textrm{1-14}$. Explicit periodic oscillation patterns are observed in $V_\textrm{3-6}$ (\textbf{a}) and $V_\textrm{3-4}$ (\textbf{b}). However, the oscillation patterns are vague in $V_\textrm{4-6}$ (\textbf{c}), $V_\textrm{2-3}$ (\textbf{d}), $V_\textrm{6-7}$ (\textbf{e}) and $V_\textrm{7-8}$ (\textbf{f}). In close inspection of \textbf{a}, \textbf{b} and \textbf{c}, we find that the patterns in $V_\textrm{3-6}$ appear to be the superposition of $V_\textrm{3-4}$ and $V_\textrm{4-6}$. And the dominant contribution to the SIP ($I_\textrm{c}$) comes from $V_\textrm{3-4}$. These observations strongly suggest that the SIP arises from a proper domain structure between terminals 3 and 4. 
}			
	\label{FigSSIPvV}
\end{figure*}


\begin{figure*}[thb]
	\renewcommand{\figurename}{Extended Data Fig.}
	\includegraphics[width=18cm]{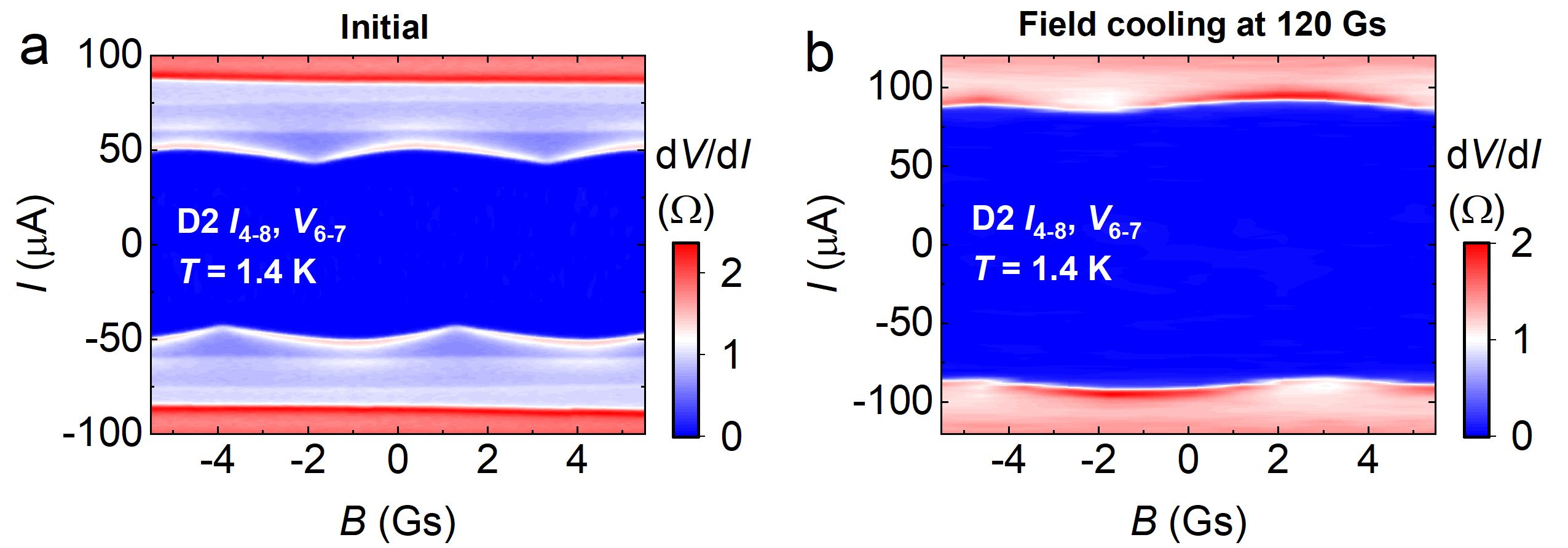}
	\caption{\textbf{Field modulation of SIPs for D2 measured at 1.4 K.} \textbf{a}, Initial SIP. \textbf{b}, SIP measured after field-cooling (FC) from $T$ slightly above $T_\mathrm{c}$ at $B= 120$~Gs. After FC, the oscillation patterns underwent significant modifications, including the alteration of period and the increase of  $|I_\mathrm{c}|$. The patterns roughly returned to their initial state after subsequent multiple thermal cycling from $T$ above $T_\textrm{c}$. This observation is distinct from what is observed for the same device but measured at different times, as shown in Extended Data Fig.~\ref{EDF10}. Relevant discussion is presented in Supplementary Note~8. 
 	}	
	\label{EDF9}
\end{figure*}

\clearpage
 \begin{figure*}[h]
	\renewcommand{\figurename}{Extended Data Fig.}
	\includegraphics[width=10cm]{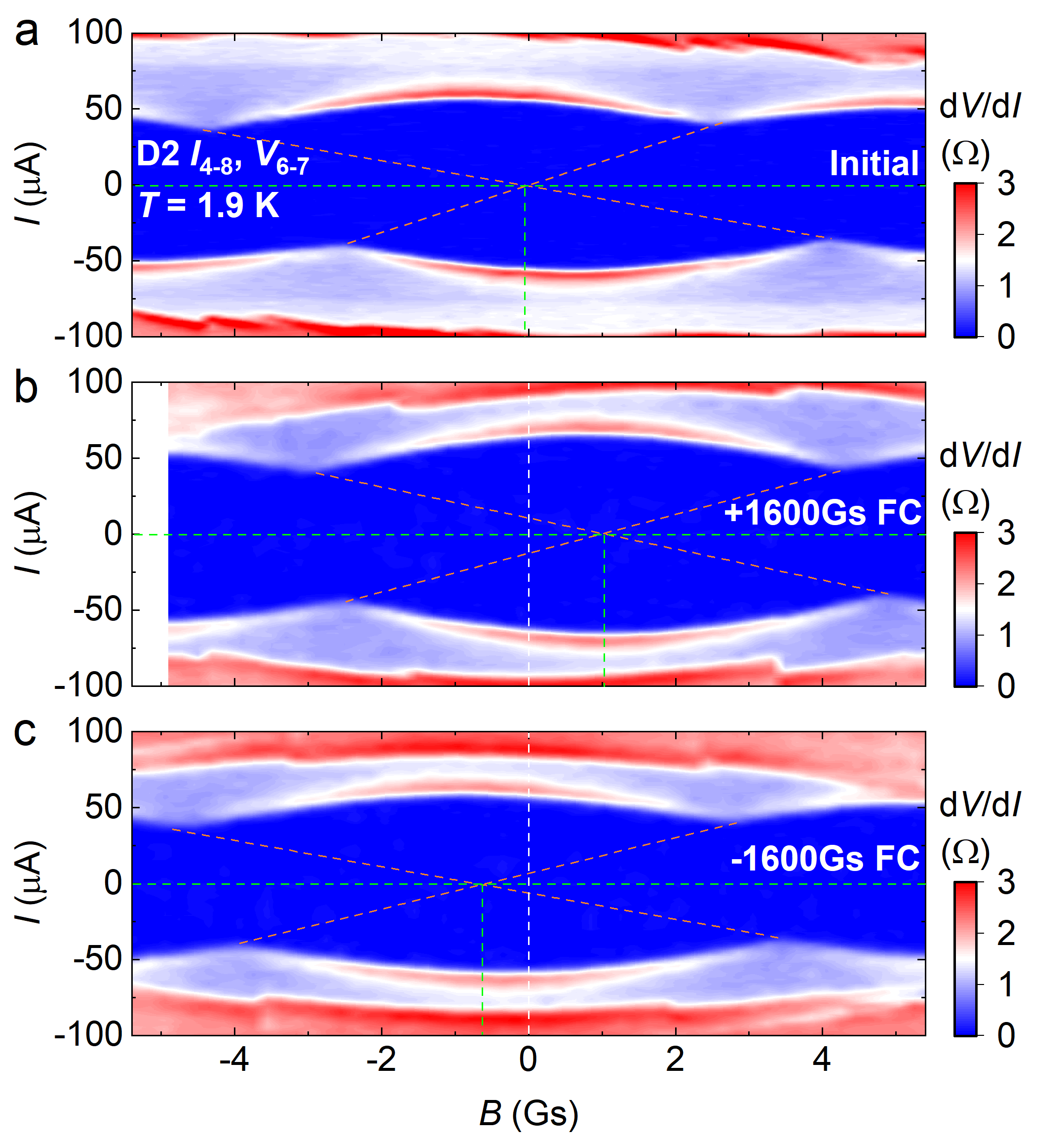}
	\caption{\textbf{Field modulation of SIPs for D2 measured at 1.9 K at different times.} \textbf{a}, Initial SIP. \textbf{b}, SIP measured after FC from $T$ slightly above $T_\mathrm{c}$ at $B=1600$~Gs. \textbf{b}, SIP measured after FC from $T$ slightly above $T_\mathrm{c}$ at $B=-1600$~Gs. Note that the period of the initial pattern is 
    distinct from that in Extended Data Fig.~\ref{EDF9}, which is due to the effect of thermal cycling from 300 K, as referring to Fig.~\ref{Fig4}d. Remarkably, the oscillation pattern remains nearly unchanged after FC with $B$ even up to 1600 Gs, a feature distinct from that observed in Extended Data Fig.~\ref{EDF9}. On closer inspection, we discern a peculiar 'advanced' nature of hysteresis in comparing the patterns among the initial, FC 1600 Gs, and FC -1600 Gs curves, i.e. the zero flux line (vertical green dashed line) shifts to positive (negative) $B$ after FC at positive (negative) field. A comprehensive discussion in Supplementary Note~8 argues that the remarkable features in Extended Data Fig.~\ref{EDF9}, ~\ref{EDF10} potentially result from chiral SC domain dynamics in response to external $B$. 
   }
   	\label{EDF10}
\end{figure*}

\end{appendix}

\end{document}